\documentclass[conference]{IEEEtran}
\IEEEoverridecommandlockouts
\usepackage{cite}
\usepackage[T1]{fontenc}
\usepackage{amsmath,amssymb,amsfonts}
\usepackage{algorithmic}
\usepackage{graphicx}
\usepackage{subcaption}
\usepackage{textcomp}
\usepackage{xcolor}
\usepackage{lipsum}
\usepackage{cases}
\usepackage{multirow}
\usepackage{comment}
\def\BibTeX{{\rm B\kern-.05em{\sc i\kern-.025em b}\kern-.08em
    T\kern-.1667em\lower.7ex\hbox{E}\kern-.125emX}}
    
\usepackage{ulem}

\begin{document}

\title{Improved feature extraction for CRNN-based multiple sound source localization}

\author{\IEEEauthorblockN{Pierre-Amaury Grumiaux}
\IEEEauthorblockA{\textit{Orange Labs}\\
Cesson-S{\'e}vign{\'e}, France \\
pierreamaury.grumiaux@orange.com}
\and
\IEEEauthorblockN{Sr\dj{}an Kiti\'c}
\IEEEauthorblockA{
\textit{Orange Labs}\\
Cesson-S{\'e}vign{\'e}, France \\
srdan.kitic@orange.com}
\and
\IEEEauthorblockN{Laurent Girin}
\IEEEauthorblockA{
\textit{Univ. Grenoble Alpes, GIPSA-lab}\\
\textit{Grenoble-INP, CNRS}\\
Grenoble, France \\
laurent.girin@grenoble-inp.fr}
\and
\IEEEauthorblockN{Alexandre Gu{\'e}rin}
\IEEEauthorblockA{
\textit{Orange Labs}\\
Cesson-S{\'e}vign{\'e}, France \\
alexandre.guerin@orange.com}
}

\vspace{-1cm}

\maketitle

\begin{abstract}

In this work, we propose to extend a state-of-the-art multi-source localization system based on a convolutional recurrent neural network and Ambisonics signals. We significantly improve the performance of the baseline network by changing the layout between convolutional and pooling layers. We propose several configurations with more convolutional layers and smaller pooling sizes in-between, so that less information is lost across the layers, leading to a better feature extraction. In parallel, we test the system's ability to localize up to 3 sources, in which case the improved feature extraction provides the most significant boost in accuracy. We evaluate and compare these improved configurations on synthetic and real-world data. The obtained results show a quite substantial improvement of the multiple sound source localization performance over the baseline network.

\end{abstract}

\begin{IEEEkeywords}
sound source localization, convolutional recurrent neural network, ambisonics, reverberation
\end{IEEEkeywords}

\section{Introduction}

Sound source localization (SSL) is a challenging task whose performance is crucial when used as a front-end in practical applications such as teleconferencing \cite{zhao}, source separation \cite{chazan}, speech enhancement \cite{xenaki}, speech recognition \cite{lee} or Human-robot interaction \cite{li2016reverberant}.
Classical subspace-based methods rely on the eigenvalue decomposition of the multichannel observed signal covariance matrix to extract the source signal(s) spatial information \cite{schmidt}, \cite{roy}. Another classical algorithm (SRP-PHAT) uses a beamformer to build an acoustic map with concentration of energy appearing in the direction of arrival (DOA) of a source \cite{dmochowski}. GCC-PHAT is another popular technique which estimates the time-difference of arrival (TDOA) for each pair of microphones to derive the DOA \cite{knapp}.
All these methods work well with a single source, and some of them can provide multiple source DOAs, but they are known to perform poorly in noisy and reverberant environment.

Recently, machine learning methods have greatly improved the performance of SSL systems. In particular, many methods based on deep neural networks (DNNs) have been proposed for estimating the DOA of one source \cite{vesperini, xiao, perotin2, yalta} or multiple sources \cite{perotin, chakrabarty, adavanne2}, leading to impressive SSL performance under challenging conditions. DNN-based SSL methods can differ on different aspects, starting with the DNN architecture: some methods propose to use a multi-layer perceptron (MLP) \cite{vesperini, xiao}, a convolutional neural network (CNN) \cite{chakrabarty, yalta, veradiaz}, a convolutional recurrent neural network (CRNN) \cite{adavanne2, perotin, perotin2} or an autoencoder (AE) \cite{liu}. Different types of input features have also been proposed, such as raw signal waveforms \cite{veradiaz}, features based on the short-time Fourier transform (STFT) \cite{chakrabarty, yalta}, correlation-based features \cite{vesperini, xiao, liu}, or Ambisonics features \cite{adavanne2, perotin, perotin2}. Finally, these works can be split into two categories according to the output type, i.e. classification \cite{xiao, chakrabarty, yalta, adavanne2, perotin, perotin2, liu} or regression \cite{vesperini, veradiaz}.

In this work, we propose an extension of the work in \cite{perotin}, which is based on a CRNN and can be considered as a state-of-the-art SSL system in the context of Ambisonics signals. First, we propose to simultaneously estimate the DOA of up to 3 sources, whereas in \cite{perotin}, SSL was limited to 2 sources. Second, we propose and compare several more complex architectures that are able to significantly improve the performance compared to \cite{perotin}. Although this may look trivial at first sight, it is not, at least from an experimental point of view. In fact, among the many architecture variants we tested, most did \textit{not} lead to improved performance. Thereby, we believe that reporting substantial improvement in SSL performance can be useful to the community and lead to a better understanding of how the spatial information is processed in a CRNN for SSL.

\section{Proposed method}

In this section, we successively describe the input features, the output configuration and the architecture of the neural networks that we trained and tested in the reported experiments.

\subsection{Input features}

As in \cite{perotin}, we work with the Ambisonics signal representation, derived from (true or simulated) recordings on a spherical microphone array \cite{zotter}. The Ambisonics format is well-suited to represent the spatial properties of a soundfield, and is, to some extent, agnostic to the microphone array configuration \cite{daniel}. It
 relies on the decomposition of the soundfield on the orthogonal basis of spherical harmonics. The number of retained coefficient defines the order of the representation: an Ambisonics representation of order $m$ requires a spherical microphone array outputing at least $(m+1)^2$ channels. In our experiments, we used first-order Ambisonics (FOA) ($m=1$) which has shown to provide sufficient spatial information for single- and multiple-speaker localization based on neural networks \cite{perotin, perotin2, fahim, adavanne, comminiello}. The four FOA coefficients are $W$ (order $0$ spherical harmonic), which can be seen as an omnidirectional microphone at the recording point, and $X$, $Y$ and $Z$ (order $1$ spherical harmonics) which can be seen as three orthogonal bidirectional microphones at the recording point. A plane wave, arriving from a direction given by an azimuth $\theta$ and elevation $\phi$, is encoded into FOA channels as follows:
\begin{equation}\label{eqFOA}
\begin{bmatrix} W(t,f) \\ X(t,f) \\ Y(t,f) \\ Z(t,f) \end{bmatrix} = \begin{bmatrix} 1 \\ 
\sqrt{3} \cos\theta \cos\phi \\ \sqrt{3} \sin\theta \cos\phi \\ \sqrt{3} \sin\phi \end{bmatrix} p(t,f),
\end{equation}
where $p(t,f)$, $t$ and $f$ denote the acoustic pressure, and STFT time and frequency bins, respectively. 

In this paper, as in \cite{perotin}, we use the active and reactive intensity vectors, respectively defined by \cite{jacobsen}: 
\begin{equation}\label{eq:activeIntensity}
    \boldsymbol{I}_a = \operatorname{Re}\{p(t,f)\boldsymbol{v}^*(t,f)\}, \quad
    \boldsymbol{I}_r = \operatorname{Im}\{p(t,f)\boldsymbol{v}^*(t,f)\},
\end{equation}
where $\boldsymbol{v}$ is the particle velocity. For a plane wave, this latter is given in the FOA representation by \cite{pulkki}: 
\begin{equation}
    \boldsymbol{v}(t,f) = \frac{1}{\rho_0 c \sqrt{3}} \begin{bmatrix}X(t,f) \\ Y(t,f) \\ Z(t,f)\end{bmatrix},
\end{equation}
where $\rho_0$ is the density of air and $c$ is the speed of sound in the air. The active intensity represents the energy flow in a particular spatial point, while the reactive intensity represents dissipative local energy transfers.
Noting that $p(t,f) = W(t,f)$ and disregarding the constant factor, the active and reactive intensity vectors in the FOA representation can be reformulated as (indexes $t$ and $f$ are omitted for concision):
\begin{equation}
    \boldsymbol{I}_a = \begin{bmatrix} \operatorname{Re}\{WX^*\} \\ \operatorname{Re}\{WY^*\} \\ \operatorname{Re}\{WZ^*\} \end{bmatrix},
    \quad
    \boldsymbol{I}_r = \begin{bmatrix} \operatorname{Im}\{WX^*\} \\ \operatorname{Im}\{WY^*\} \\ \operatorname{Im}\{WZ^*\} \end{bmatrix}.
\end{equation}
For each time-frequency (TF) bin, the above STFT-domain active and reactive intensity vectors are concatenated to form a $6$-channel vector. This vector is then normalized by dividing it by the sound power given by $|W(t,f)|^2 + \frac{1}{3}(|X(t,f)|^2+|Y(t,f)|^2+|Z(t,f)|^2)$, which is reminiscent of the so-called \textit{Frequency Domain Velocity Vector} representation \cite{tdvv}. Then, the normalized vectors are concatenated across time and frequency bins to form a 3D $T \times F \times 6$ input tensor, where $T$ is the number of frames and $F$ is the number of frequency bins. In our experiments, we used signals sampled at $16$~kHz, a 1,024-point ($64$ ms) STFT (hence $F = 513$) with a sinusoidal analysis window and 50\% overlap. 
Each input sequence given to our CRNN contains $T = 25$ frames (i.e.~about $800$~ms of signal), hence an input tensor is of size $25 \times 513 \times 6$.

\subsection{Output}

We choose the classification approach for our experiments, which straightforwardly allows for single or multiple sound source localization. To do that, as in \cite{perotin}, we divide the 2D unit sphere into a quasi-uniform grid. The candidate elevations $\phi_i \in [-90,90]$ and azimuths $\theta_i^j \in [-180,180]$ on the grid are given by:
\begin{equation}
    \begin{cases}
        \phi_i = -90 + \frac{i}{I} \times 180 & \text{with $i \in \{0, ..., I\}$} \\
        \theta_i^j = -180 + \frac{j}{J^i+1} \times 360 & \text{with $j \in \{0, ..., J^i\}$}, \\
    \end{cases}
\end{equation}
where $I = \lfloor \frac{180}{\alpha} \rfloor$ and $J^i = \lfloor \frac{360}{\alpha} cos \phi_i \rfloor$ with $\alpha$ the grid resolution in degrees.
Each zone around a point of the spherical grid corresponds to a class. The output target of our neural networks is represented by a vector $y$ of size $C$ with ``binary'' entries, where $C$ is the total number of classes. If a source is present in a zone corresponding to class $c$, $y(c) = 1$, otherwise $y(c) = 0$. As we address the multiple source localization problem, $y$ can contain multiple entries set to 1.

\subsection{Network architectures}

The architectures we used in the experiments reported in the present paper are extensions/variants of the CRNN proposed in \cite{perotin}, which has 3 convolutional layers, followed by a bidirectional LSTM network (BiLSTM), followed by a fully-connected layer. We aim at improving the SSL performance by modifying different aspects of this architecture. First, we propose to increase the number of convolutional layers to extract more ``high-level'' features that can be efficiently processed by the BiLSTM. In our experiments, we varied the number of convolutional blocks (each containing 2 successive convolutional layers followed by a max-pooling layer) from 4 to 7. Second, in \cite{perotin} each convolutional block is followed by a max-pooling operation with a quite large pooling size ($1 \times 8$ for blocks 1 and 2 and $1 \times 4$ for block 3). This was to ensure that the dimension at the output of the last convolutional block is significantly reduced for an efficient processing by the BiLSTM block. However, such a large pooling size can cause an important loss of information through the convolutional layers. In order to alleviate this problem, we propose to exploit our larger number of convolutional layers to i) apply max-pooling only every 2 convolutional layers, and ii) reduce the pooling size (details are given in Section~\ref{subsec:config}).

\begin{figure}
   \begin{subfigure}[c]{0.46\columnwidth}
        \includegraphics[width=\linewidth]{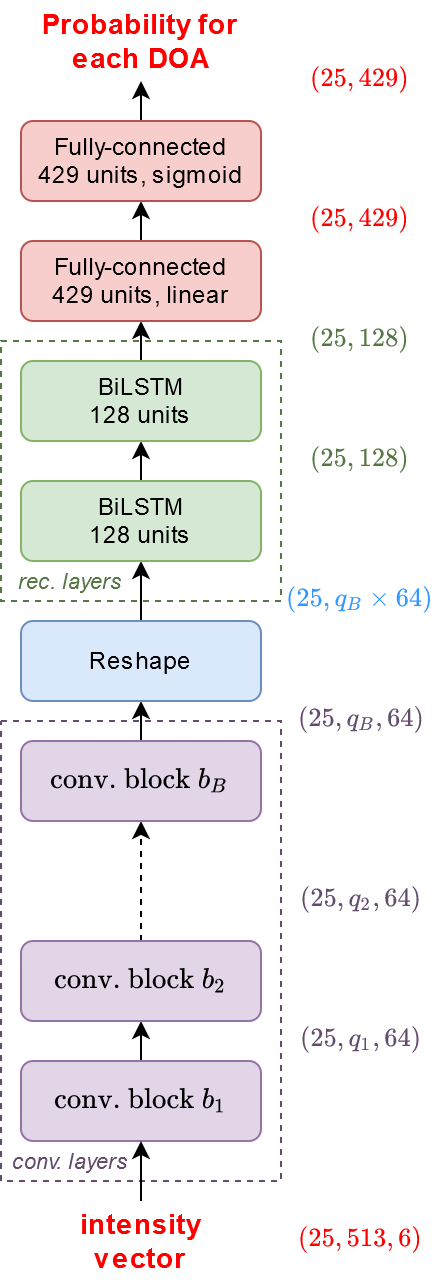}
    \end{subfigure}
    \begin{subfigure}[c]{0.54\columnwidth}
        \includegraphics[width=\linewidth]{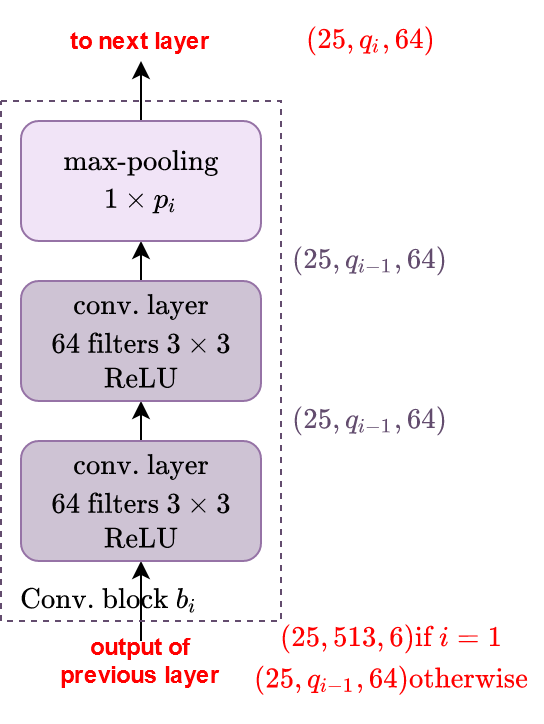}
    \end{subfigure}
    \caption{The proposed improved CRNN architecture. Left: General architecture; Right: detail of one convolutional block. The dimension of the input/output data for each layer is indicated on the right in between layers.}
    \label{fig:genericArchi}
\end{figure}

Fig.~\ref{fig:genericArchi} shows the generic architecture of the CRNNs we used in the reported experiments. It is composed of $B$ convolutional blocks, $B \in \{4, 5, 6, 7\}$, followed by two BiLSTM layers, and ends with two fully-connected layers. Each convolutional block $b_i$ ($i \in [1, ..., B]$) is composed of 2 convolutional layers with $64$ filters of size $3 \times 3$, then a max-pooling layer of size $1 \times P_i$, where $P_i$ values are varied in our experiments. Unlike in \cite{perotin}, where a max-pooling layer is used after each convolutional layers, we use only one max-pooling layer after two convolutional layers to allow the network to extract more high-level features before downsampling. Note that if $q_i$ denotes the second dimension of the output of block $b_i$, we have: $q_i = \frac{q_{i-1}}{p_i}$.

\section{Experiments}

\subsection{Data}

Our training dataset was generated using synthetic spatial room impulse responses (SRIRs), in the same way as in \cite{perotin}. We adapted the SRIR simulator \cite{rirGenerator} (based on the image-source method \cite{imageSource}), such that it yields the FOA impulse responses. We (randomly) generated many configurations,  by varying the size and shape of ``shoebox rooms,'' the reverberation time (RT60), and the source and microphone positions. Microphone signals were obtained by convolving TIMIT speech signals \cite{timit} with the simulated SRIRs. For each room configuration, we generated the signals corresponding to a single source, 2- and 3-source mixtures, all 1~s long. 
Diffuse babble noise was added to those mixtures with a random SNR between 0 and 20~dB. These 1s-long mixtures are finally transformed into 2 sequences of 25 frames with 50\% frame overlap. Finally, for our training dataset, we end up with 257,400 sequences for each number of speakers considered in the experiments, that is 1 to 3 speakers, leading to a total of 772,200 training sequences (about 172 hours).

To test our models, we used three types of datasets: i) a dataset based on synthetic SRIRs, that was generated in the same way as for the training but with different SRIRs, speech and noise signals; ii) a dataset based on recorded SRIRs in our acoustic lab (RT60 $\approx$ 500~ms), using all possible combinations from 36 microphone positions and 16 loudspeaker positions; iii) the real-world evaluation dataset from the LOCATA challenge \cite{locata}, with raw Eigenmike recordings converted into the FOA format. We considered LOCATA Task 1 (single static source), Task 2 (multiple static sources), Task 3 (single moving source) and Task 4 (multiple moving sources).

\subsection{Configurations}
\label{subsec:config}

\begin{table}[t]
\resizebox{\columnwidth}{!}{%
\begin{tabular}{|c|c|c|c|c|c|c|c|c|}
\hline
\textbf{Config.} & \textbf{\# parameters} & \textbf{$P_1$} & \textbf{$P_2$} & \textbf{$P_3$} & \textbf{$P_4$} & \textbf{$P_5$} & \textbf{$P_6$} & \textbf{$P_7$} \\\hline
4-2 & 700,259 & 8 & 4 & 4 & 2 & - & - & - \\
4-4 & 765,795 & 4 & 4 & 4 & 2 & - & - & - \\
4-8 & 896,867 & 4 & 4 & 2 & 2 & - & - & - \\
5-2 & 774,315 & 4 & 4 & 4 & 2 & 2 & - & - \\
5-4 & 839,851 & 4 & 4 & 2 & 2 & 2 & - & - \\
6-2 & 848,371 & 4 & 4 & 2 & 2 & 2 & 2 & - \\
6-4 & 913,907 & 4 & 2 & 2 & 2 & 2 & 2 & - \\
7-2 & 922,427 & 4 & 2 & 2 & 2 & 2 & 2 & 2 \\
7-4 & 987,963 & 2 & 2 & 2 & 2 & 2 & 2 & 2 \\
\hline
\end{tabular}}
\caption{Max-pooling sizes  $P_i$ of the successive convolutional blocks $b_i$ of the improved CRNN. The corresponding total number of parameters is given in column 2.}
\label{tab:archiParams}
\end{table}

Table~\ref{tab:archiParams} details the CRNN configurations used in our experiments. As stated before, the configurations differ in the number of convolutional blocks $B$ and the max-pooling size $P_i$ of each convolutional block $b_i$. The tested combinations of $P_i$ values are such that the size of the data second dimension after the last max-pooling layer $q_B$ is equal to $2$, $4$ or $8$. For concision, a configuration name is of the form ``$B$-$q_B$'', e.g.~$4$-$2$ stands for $B=4$ and $q_B=2$.

\begin{table*}[h!]
\centering
\resizebox{2.\columnwidth}{!}{%
\begin{tabular}{|c|cccc|cccc|cccc|}
\hline
\multirow{2}{*}{\textbf{Model}} & \multicolumn{4}{c|}{\textbf{1 source}}                                                      & \multicolumn{4}{c|}{\textbf{2 sources}}                                                     & \multicolumn{4}{c|}{\textbf{3 sources}}                                                     \\
                                & \textbf{Acc. \textless{}10°} & \textbf{Acc. \textless{}15°} & \textbf{Mean} & \textbf{Med.} & \textbf{Acc. \textless{}10°} & \textbf{Acc. \textless{}15°} & \textbf{Mean} & \textbf{Med.} & \textbf{Acc. \textless{}10°} & \textbf{Acc. \textless{}15°} & \textbf{Mean} & \textbf{Med.} \\ \hline
\multicolumn{1}{|l|}{Baseline \cite{perotin}} & 94.6                         & 99.2                         & 5.2           & 4.7           & 77.6                         & 85.7                         & 15.3          & 6.0           & 57.1                         & 68.1                         & 27.2          & 8.3           \\
4-2                        & 97.6                         & 99.6                         & 4.7           & 4.2           & 86.7                         & 92.5                         & 8.3          & 4.9           & 71.4                         & 79.9                         & 15.0          & 6.3           \\
4-4                        & 98.3                         & \textbf{99.7}                         & 4.5           & \textbf{4.1}           & 87.9                         & 92.7                         & 8.5          & 4.8           & 71.2                         & 79.5                         & 15.4          & 6.2           \\
4-8                        & 98.2                         & 99.6                         & 4.5           & \textbf{4.1}           & 88.0                         & 92.6                         & 8.4          & 4.8           & 72.2                         & 80.7                         & 14.7          & 6.1           \\
5-2                        & 98.3                         & \textbf{99.7}                         & 4.6           & \textbf{4.1}           & 87.9                         & 92.8                         & 8.0          & 4.7           & 72.5                         & 80.5                         & 14.6          & 6.1           \\
5-4                        & 98.4                         & \textbf{99.7}                         & 4.6           & \textbf{4.1}           & 88.8                         & 93.1                         & 8.1          & 4.7           & 73.3                         & 81.0                         & 14.9          & 6.1           \\
6-2                        & 98.4                         & 99.5                         & 4.7           & \textbf{4.1}           & 88.7                         & 93.2                         & 8.1          & 4.8           & 72.2                         & 80.7                         & 14.4          & 6.1           \\
6-4                        & \textbf{98.6}                         & \textbf{99.7}                         & \textbf{4.4}           & \textbf{4.1}           & 88.3                         & 93.3                         & \textbf{7.7}          & 4.7           & \textbf{74.7}                         & \textbf{83.4}                         & \textbf{12.8}          & 5.9           \\
7-2                        & 97.8                         & 99.5                         & 4.7           & \textbf{4.1}           & 86.3                         & 92.0                         & 8.6          & 4.8           & 68.6                         & 77.4                         & 16.3          & 6.5           \\
7-4                        & 98.4                         & \textbf{99.7}                         & \textbf{4.4}           & \textbf{4.1}           & \textbf{89.2}                         & \textbf{93.5}                         & 7.8          & \textbf{4.6}           & 74.4                         & 81.3                         & 14.1          & \textbf{5.8}           \\ \hline
\end{tabular}}
\caption{SSL results on the test dataset generated with synthetic SRIRs (best results are in bold).}
\label{tab:resultsSynthetic}
\end{table*}

\begin{table*}[h!]
\centering
\resizebox{2.\columnwidth}{!}{%
\begin{tabular}{|c|cccc|cccc|cccc|}
\hline
\multirow{2}{*}{\textbf{Model}} & \multicolumn{4}{c|}{\textbf{1 source}}                                                      & \multicolumn{4}{c|}{\textbf{2 sources}}                                                     & \multicolumn{4}{c|}{\textbf{3 sources}}                                                     \\
                                & \textbf{Acc. \textless{}10°} & \textbf{Acc. \textless{}15°} & \textbf{Mean} & \textbf{Med.} & \textbf{Acc. \textless{}10°} & \textbf{Acc. \textless{}15°} & \textbf{Mean} & \textbf{Med.} & \textbf{Acc. \textless{}10°} & \textbf{Acc. \textless{}15°} & \textbf{Mean} & \textbf{Med.} \\ \hline
\multicolumn{1}{|l|}{Baseline \cite{perotin}} & 75.2                         & 91.9                         & 8.3          & 6.3           & 59.8                         & 75.2                         & 16.7          & 8.3           & 44.3                         & 58.4                         & 26.2          & 11.9          \\
4-2                        & 77.7                         & 92.9                        & 8.1           & \textbf{6.1}           & 67.5                         & 83.6                         & 12.9          & 7.4           & 53.3                         & 67.5                         & 21.3          & 9.2           \\
4-4                        & 77.7                         & 93.2                         & 7.9           & \textbf{6.1}           & 66.7                         & 83.4                         & 12.9          & 7.5           & 54.0                         & 69.2                         & 20.4          & 9.1           \\
4-8                        & 77.8                         & 92.7                         & 8.1           & \textbf{6.1}           & 69.2                         & 84.1                         & 12.5          & 7.2           & 54.8                         & 69.7                        & 20.5          & 9.1           \\
5-2                        & 77.0                         & 93.6                         & 7.9           & 6.2           & 68.1                         & 84.1                         & 12.1          & 7.3           & 55.3                         & 69.6                         & 18.7          & 8.9           \\
5-4                        & 78.7                         & \textbf{93.8}                         & \textbf{7.6 }         & 6.2           & \textbf{70.2}                        & 86.0                         & 12.0          & \textbf{7.0 }         & 55.4                        & 70.9                        & 19.7          & 8.9           \\
6-2                        & 78.1                         & 93.6                       & \textbf{7.6}           & \textbf{6.1}           & 68.5                         & 84.8                         & 11.9          & 7.1           & 53.9                         & 69.3                         & 19.7          & 9.1           \\
6-4                        & 79.0                         & 93.7                         & \textbf{7.6}           & \textbf{6.1}           & 68.2                         & 84.7                         & 11.9          & 7.2           & 56.8                        & \textbf{73.3}                         & \textbf{17.3}          & 8.7           \\
7-2                        & 76.6                         & 93.4                         & 7.7           & 6.3           & 68.0                         & 83.7                        & 12.2          & 7.2           & 53.3                         & 66.8                         & 20.9          & 9.3           \\
7-4                        & \textbf{79.8}                         & 93.6                         & 7.7           & \textbf{6.1}           & 68.6                         &\textbf{86.2}                         & \textbf{11.7}          & 7.2           & \textbf{56.9}                         & 70.7                         & 19.6          & \textbf{8.6}           \\ \hline
\end{tabular}}
\caption{SSL results on the dataset generated with real SRIRs (best results are in bold).}
\label{tab:resultsRealSrir}
\end{table*}

\begin{table*}[h!]
    \centering
    \resizebox{2.\columnwidth}{!}{%
    \begin{tabular}{|c|ccccc|ccccc|ccccc|ccccc|}
    \hline
    \multirow{3}{*}{\textbf{Model}} & \multicolumn{5}{c|}{\textbf{Task 1}}                                                                                                             & \multicolumn{5}{c|}{\textbf{Task 2}}                                                                                                            & \multicolumn{5}{c|}{\textbf{Task 3}}                                                                                                            & \multicolumn{5}{c|}{\textbf{Task 4}}                                                                                                            \\ \cline{2-21}
                                    & \multicolumn{3}{c}{\textbf{Accuracy}}                                       & \multirow{2}{*}{\textbf{Mean}} & \multirow{2}{*}{\textbf{Median}} & \multicolumn{3}{c}{\textbf{Accuracy}}                                      & \multirow{2}{*}{\textbf{Mean}} & \multirow{2}{*}{\textbf{Median}} & \multicolumn{3}{c}{\textbf{Accuracy}}                                      & \multirow{2}{*}{\textbf{Mean}} & \multirow{2}{*}{\textbf{Median}} & \multicolumn{3}{c}{\textbf{Accuracy}}                                      & \multirow{2}{*}{\textbf{Mean}} & \multirow{2}{*}{\textbf{Median}} \\
                                    & \textbf{\textless{}10°} & \textbf{\textless{}15°} & \textbf{\textless{}20°} &                                &                                  & \textbf{\textless{}10°} & \textbf{\textless{}15°} & \textbf{\textless{}20°} &                                &                                  & \textbf{\textless{}10°} & \textbf{\textless{}15°} & \textbf{\textless{}20°} &                                &                                  & \textbf{\textless{}10°} & \textbf{\textless{}15°} & \textbf{\textless{}20°} &                                &                                  \\ \hline
    Baseline \cite{perotin}      &35.8                    & 50.7                   & 96.6                & 13.5     & 14.9     & 30.5                   &68.2                    &  91.8                  & 13.9     & 14.1           & 29.5                  &  66.9             & 86.9           & 14.1              & 12.7                      & 30.3                   & 56.7                 & 85.6                   & 15.1     & 14.1       \\
    4-2       &40.0                    & 51.7                   & 99.2                   & 13.0     & 13.0     & 26.9                   & 71.7                   & 91.9                   & 14.0     & 12.9           & 36.0                  &  71.2             & 88.5           & 13.4              & 12.0     	                & 55.2                   & 73.1                 &  89.9                  &12.5      & 8.6         \\
    4-4       & 40.9                   & 50.6                   & 98.8                   &13.3      &13.0      & \textbf{34.5}          & 69.7                   & 91.6                   & 13.4     & 13.0           & 40.1                  &  70.0             & 88.4           & 12.4              &  11.8                     & 53.9                   & 71.0                   &  91.2                  & 12.2     & 8.4       \\
    4-8       & 36.4                   &51.5                    & 91.0                   &13.5      & 14.9     & 29.1                   & 73.3                   & 91.5                   & 14.0     & 13.0           & 41.3                  & 72.3              & 91.7           & 12.0              & 11.5                      & 52.2                   & 70.9                   & 90.8                   & 11.9     & 9.3       \\
    5-2       & 39.2                   &  49.7                  & 99.0                   & 13.5     & 16.1     & 27.0                   & 70.1                   & 93.1                   &13.6      & 13.6           & 40.2                  & 72.5              & 91.7           & 11.9              & 11.7                      & 52.0                   & 70.1                   & 91.6                   & 11.8     &9.2        \\
    5-4       & \textbf{46.5}          &   51.2                 & 98.4          & \textbf{12.4}     & \textbf{12.0}     & 32.8          & \textbf{73.4}          & 92.5                   & 13.4     & \textbf{12.5 } & 45.3                  & 77.7              & 91.3           & 11.8              &  \textbf{10.8}            & 56.2                   & 73.6                  & 91.7                   & 11.7     & 8.1        \\
    6-2       & 35.3                   &  52.5                  & 95.8                   &13.7      & 14.5     & 22.4                   &   67.4                 & 90.4                   & 14.6     & 13.7           & 35.8                  & 71.7              & 90.9           & 13.7              & 12.1                      & 49.7                   & 68.1                   & 89.2                  & 12.7     &10.1        \\
    6-4       & 23.5                   &  49.4                  & 98.3                   & 14.4     &15.8      & 19.1                   & 66.5                   &92.4                    & 14.5     & 13.7           & 28.7                  & 62.6              & 88.4           & 13.7              &  13.3                     & 38.2                   & 59.8                   & 88.3                   & 13.8     & 13.6     \\
    7-2       &  32.9                  &  \textbf{55.5}         & 91.6                   & 16.0     & 13.0     & 23.2                   & 69.3                   &91.3                   & 14.7     & 13.6            & 35.8                  & 71.0              & 86.9           & 15.7              & 12.2                      & 48.8                   & 69.0                   & 89.3                   & 12.8     & 10.3     \\
    7-4       &  43.6                  &  49.4                  & \textbf{99.4 }         & 12.7     & 16.1     & 32.0                   & 71.5              & \textbf{93.8}          & \textbf{13.0}  & 13.0          & \textbf{45.6}         & \textbf{75.3}     & \textbf{92.5}  & \textbf{11.3}     & \textbf{10.8}             & \textbf{58.8}          & \textbf{74.0}          & \textbf{92.0}       & \textbf{11.1}     & \textbf{7.6}  \\ \hline
    \end{tabular}}
\caption{SSL results on the LOCATA evaluation dataset (best results are in bold).}
\label{tab:resultsLocata}
\end{table*}

\subsection{Training procedure}

The CRNNs were designed with Keras and trained using the Nadam optimizer \cite{nadam} with default parameters and Nvidia GTX1080 GPUs. Early stopping was applied with a patience of 20 epochs and the learning rate was divided by 2 with a patience of 10 epochs, both by monitoring the accuracy on the validation set. The maximum number of epochs was 300. 

\subsection{Evaluation procedure}

When inferring the DOAs in test examples, we average the output of the trained network in the frame dimension, i.e. we obtain one probability value per DOA for a 25-frame sequence. Unlike in \cite{perotin}, we do not smooth the probability distribution within a neighborhood since we found out that the results were degraded compared to using the raw distribution. Instead, we directly keep the $S$ highest peaks where $S$ is the number of sources (a peak represents the local maximum of probability distribution within the spherical geometry). In the first two experiments, we suppose that $S$ in known, while for the experiments based on the LOCATA dataset we use the fixed threshold $\beta=0.2$ to detect source(s).

\subsection{Metrics and baselines}

We evaluated the DOA estimation in terms of sequence-wise accuracy, i.e. the percentage of 25-frame sequences whose DOA yields to an angular error less than a certain tolerance. Here, the tolerance we used was either 10\textdegree~or 15\textdegree, considering the fact that the minimum angle between two points in our grid is 7\textdegree. We also evaluated the mean and median angular error, averaged over all test sequences. 
We compared the proposed improved CRNN (with different configuration settings) and the CRNN of \cite{perotin} which, again, can be considered as a state-of-the-art baseline (indeed, this baseline network was shown in \cite{perotin} to perform better than the algorithm proposed in \cite{TRAMP}, with the latter known for outperforming the LOCATA baseline \cite{locata}). Note that in \cite{perotin}, training and test is made for up to 2 speakers. For fair comparison, we trained their network on our dataset with up to 3 speakers.

\subsection{Results}

Tables~\ref{tab:resultsSynthetic}  and \ref{tab:resultsRealSrir} show the performance of all the tested architectures (including the baseline CRNN) on the dataset generated with synthetic SRIRs and the one generated with real SRIRs, respectively. In a general manner, we observe that the new architectures provide a substantial improvement in performance over the baseline network, with the best performance obtained by Models 6-4 and 7-4, and with a larger improvement for multi-source localization than for 1-source localization. Indeed, for 1-source signals, Model 6-4 reaches for example an accuracy of 98.6\% (for 10° tolerance) with a mean angular error of 4.4° on data generated with synthetic SRIRs, vs 94.6\% and 5.2° for the baseline CRNN (which seems already well-tuned for monosource localization). For 2-source signals, the accuracy of Model 7-4 is 89.2\%, that is 11.6\% over the baseline, and its mean angle error is 7.8°, that is 7.5° better than baseline. For 3-source signals (and synthetic SRIRs), Model 6-4 performs at 74.7\% accuracy, which is 17.6\% over the baseline, and 12.8° mean angular error, which is 14.4° better than the baseline. The relative improvement over the baseline is thus here about 31\% (in accuracy) and more than 53\% (in mean angle error), which is a quite substantial gain in performance. 
The same tendency can be observed looking at the accuracy for an angle tolerance of 15° and/or for the real SRIRs. In short, for multi-source localization, we can clearly see that the proposed use of more convolutional blocks with less pooling allows a better feature extraction, resulting in a more robust multi-source localization, with less outliers (\textit{i.e.} with less estimates being very far from the ground-truth).

The results obtained on the LOCATA dataset are given in Table~\ref{tab:resultsLocata}. While we notice a classical drop in performance for all tested models compared to synthetic data, these results confirm the trend observed in the previous two experiments: Here also, the proposed new architectures yield a substantial improvement over the baseline for all metrics. Interestingly, the largest gains in accuracy (with respect to the baseline), as well as in the reduction of mean and median angular errors, are observed for LOCATA Tasks 3 and 4 which consider mobile sources (the best results are obtained with the 7-4 model).

In a general manner, the largest tested model (7-4) performs best in our experiments with respect to several metrics, but the results for smaller architectures such as 6-4 are quite close (and sometimes even better). In fact, Model 6-4 systematically performs better than Model 7-2. We thus see a tendency that the higher the $q_B$ dimension, the better the results, and a larger $q_B$ value can compensate for a lower number of convolutive blocks. 

\section{Conclusion}

We have modified the feature extraction part of a state-of-the-art CRNN-based source localization neural network, leading to substantial performance improvements, as observed in new experiments using synthetic and real-world data (up to 30\% relative gain in accuracy and 52\% in mean angle error). While the improvement in the single-source scenario is incremental, the novel architectures largely outperform the baseline in the more challenging multi-source setting. Future work will focus on structural modifications of the remaining parts of the CRNN architecture, namely by adapting the recurrent, and/or the network output layers.

\end{document}